\documentclass[prd,superscriptaddress,amsfonts,amssymb,amsmath,showpacs,onecolumn]{revtex4-2}
\usepackage{bm}
\usepackage{amsfonts}
\usepackage{graphicx}
\usepackage{amsmath}
\usepackage{palatino}
\usepackage{mathpazo}
\usepackage{textcomp}
\usepackage[utf8]{inputenc}
\usepackage[T1]{fontenc}
\usepackage{float}
\usepackage{booktabs}
\usepackage{dcolumn}
\usepackage{multirow}
\usepackage{hyperref}
\hypersetup{colorlinks,citecolor=blue}
\usepackage{amsmath}
\usepackage{xcolor}
\usepackage{orcidlink}
\usepackage[caption=false]{subfig}
\usepackage{commath}
\usepackage{amssymb}

\usepackage{esdiff}
\usepackage{mathtools}
\usepackage{ifthen}
\usepackage{mathtools, amssymb} 

\usepackage{siunitx}
\usepackage{tikz}
\usetikzlibrary{hobby} 
\usetikzlibrary{arrows.meta} 
\tikzset{>={Latex[length=4,width=4]}} 
\usetikzlibrary{calc,intersections,decorations.markings}
\usepackage{siunitx}
\usepackage{xcolor} 

\colorlet{mylightblue}{blue!20}
\colorlet{myblue}{blue!50!black}
\colorlet{mydarkblue}{blue!30!black}
\colorlet{mylightred}{red!10}
\colorlet{myred}{red!50!black}
\colorlet{mydarkred}{red!60!black}
\colorlet{mydarkgreen}{green!30!black}

\tikzset{
  midarr/.style={decoration={markings,mark=at position #1 with {\arrow{stealth}}},postaction={decorate}},
  midarr/.default=0.5
}

\def\jnl@style{\it}
\def\aaref@jnl#1{{\jnl@style#1}}

\def\aaref@jnl#1{{\jnl@style#1}}

\def\aj{\aaref@jnl{AJ}}                   
\def\apj{\aaref@jnl{ApJ}}                 
\def\apjl{\aaref@jnl{ApJ}}                
\def\apjs{\aaref@jnl{ApJS}}               
\def\apss{\aaref@jnl{Ap\&SS}}             
\def\aap{\aaref@jnl{A\&A}}                
\def\aapr{\aaref@jnl{A\&A~Rev.}}          
\def\aaps{\aaref@jnl{A\&AS}}              
\def\mnras{\aaref@jnl{Mon.~Not.~Roy.~Astron.~Soc.}}             
\def\prd{\aaref@jnl{Phys.~Rev.~D}}        
\def\prc{\aaref@jnl{Phys.~Rev.~C}}  
\def\prl{\aaref@jnl{Phys.~Rev.~Lett.}}    
\def\qjras{\aaref@jnl{QJRAS}}             
\def\skytel{\aaref@jnl{S\&T}}             
\def\ssr{\aaref@jnl{Space~Sci.~Rev.}}     
\def\zap{\aaref@jnl{ZAp}}                 
\def\nat{\aaref@jnl{Nature}}              
\def\aplett{\aaref@jnl{Astrophys.~Lett.}} 
\def\apspr{\aaref@jnl{Astrophys.~Space~Phys.~Res.}} 
\def\physrep{\aaref@jnl{Phys.~Rep.}}      
\def\physscr{\aaref@jnl{Phys.~Scr}}       
\def\commat{\aaref@jnl{Comm.~Math.~Phys.}}              
\def\science{\aaref@jnl{Science}}               
\def\cqg{\aaref@jnl{Classical Quant.~Grav.}}            
\def\jpcs{\aaref@jnl{JPCS}}                                     
\def\ijmpd{\aaref@jnl{Int.~J.~Mod.~Phys.~D}}                    
\def\grg{\aaref@jnl{Gen.~Relat.~Gravit.}}               
\def\rpp{\aaref@jnl{Rep.~Prog.~Phys.}}          
\def\npa{\aaref@jnl{Nucl.~Phys.~A}}        
\def\lrr{\aaref@jnl{Living Rev.~Rel.}}                   
\def\jcap{\aaref@jnl{J.~Cosmology Astropart.~Phys.}}    
\def\rmp{\aaref@jnl{Rev.~Mod.~Phys.}}   
\def\epjc{\aaref@jnl{Eur.~Phys.~J.~C}}

\hypersetup{linkcolor=blue}

\allowdisplaybreaks[1]

\begin{document}

\color{black}       

\title{Bianchi type-I Barrow holographic dark energy model in
symmetric teleparallel gravity}

\author{M. Koussour\orcidlink{0000-0002-4188-0572}}
\email{pr.mouhssine@gmail.com}
\affiliation{Quantum Physics and Magnetism Team, LPMC, Faculty of Science Ben
M'sik,\\
Casablanca Hassan II University,
Morocco.} 

\author{S.H. Shekh\orcidlink{0000-0003-4545-1975}}
\email{da\_salim@rediff.com}
\affiliation{Department of Mathematics. S. P. M. Science and Gilani Arts Commerce
College,\\ Ghatanji, Dist. Yavatmal, Maharashtra-445301, India.}

\author{M. Bennai\orcidlink{0000-0002-7364-5171}}
\email{mdbennai@yahoo.fr}
\affiliation{Quantum Physics and Magnetism Team, LPMC, Faculty of Science Ben
M'sik,\\
Casablanca Hassan II University,
Morocco.} 
\affiliation{Lab of High Energy Physics, Modeling and Simulations, Faculty of
Science,\\
University Mohammed V-Agdal, Rabat, Morocco.}

\date{\today}
\begin{abstract}

In this work, we have discussed a spatially homogeneous
and anisotropic Bianchi type-I space-time in the presence of Barrow
holographic dark energy (infrared cut-off is the Hubble's horizon) proposed
by Barrow recently (Physics Letters B 808 (2020): 135643) and matter in the
framework of $f(Q)$ gravity where the non-metricity $Q$ is responsible for
the gravitational interaction for the specific choice of $f(Q)=\lambda Q^{2}$
(where $\lambda <0$ is a constant). To find the exact solutions to the field
equations we consider the deceleration parameter $q$ is a function of the
Hubble's parameter $H$ i.e. $q=b-\frac{n}{H}$ (where $b$ and $n$ are
constants). We have studied the physical behavior of important cosmological
parameters such as the EoS parameter, BHDE and matter density, skewness
parameter, squared sound speed, and $\omega _{B}-\omega _{B}^{^{\prime }}$
plane. Also, we constrain the values of the model parameters $b$ and $n$
using $57$ Hubble's parameter measurements.

\end{abstract}

\maketitle
\section{Introduction}

\label{sec1}

The astrophysical clarifications established the occurrence of dark energy
in the Cosmos and the modification in its stake in the mass energy budget of
the Cosmos. Dark energy is supposed to be accountable for the speeded up
opening out of the Cosmos \cite{ref1, ref2, ref3, ref4}. The dark energy
produce a negative equation of state parameter which is the ratio of the
pressure $(p)$ to its energy density $(\rho )$ as $(\omega =\frac{p}{\rho })$%
. The modest dark energy candidate is the vacuum energy $(\omega =-1)$ which
is accurately analogous to the cosmological constant $(\Lambda )$. The
supplementary conventional development of action, which can be designated by
minimally coupled scalar fields, such are quintessence field $(\omega >-1)$,
phantom field $(\omega <-1)$ and quintom (that can across from phantom
region to quintessence region as evolved) and $\omega \ll -1$ is fixed just
before the present cosmological data also have time-dependent equation of
state parameter. The observational results that come from Supernovae type-Ia
(SNe-Ia) data and a mixture of SNe-Ia data with Cosmic Microwave Background
Radiation (CMBR) and GRS respectively acquired $-1.66<\omega <-0.62$ and $%
-1.33<\omega <-0.79$. Further, the most modern conclusion in 2009 reserved
from the dark energy equation of state parameter to $-1.44<\omega <-0.92$.
The present Planck data tells that there is $68.3\%$ dark energy of the
total energy contents of the universe. A variety of dark energy models have
been deliberated in which cosmological constant is the crucial dark energy
candidate for re-counting dark energy phenomenon but it has some thoughtful
problems. Due to some serious problems, several dark energy models like a
family of scalar fields such as quintessence, phantom, quintom, tachyon,
K-essence along with various Chaplygin gas models like generalized Chaplygin
gas, extended Chaplygin gas and modified Chaplygin gas has been produced.

In the middle of altered dynamical dark energy models, particularly, the
holographic dark energy model has been turn out to be a positive way as of
late to anticipate the dark energy mystery. It is observed that holographic
dark energy are much helpful to explain the accelerated expansion of the
cosmos \cite{ref5, ref6, ref7, ref8}. Some of the models of Holographic dark
energy are mentioned in \cite{ref9, ref10, ref11, ref12, A1, V2}. Next, the
holographic dark energy model by changing standard holographic dark energy
as Barrow holographic dark energy. Barrow holographic dark energy model is
an inventive one surrounded by the existing approach in enlightening the
recent speeding up of the universe. Bearing in mind quantum gravitational
effects, Barrow have articulated, the black hole surface. Succeeding this
viewpoint, Barrow have suggested that the black hole entropy would gratify a
more universal relation \cite{ref13}, 
\begin{equation}
S_{h}=\left( \frac{A}{A_{0}}\right) ^{(1+\bigtriangleup /2)},  \label{e1}
\end{equation}%
where $A$ and $A_{0}$ are the area of the black hole horizon and plank area
respectively. $\bigtriangleup $ be the exponent which computes the extent of
quantum-gravitational deformation effects.

The standard holographic dark energy is given by the inequality $\rho
L^{4}\leq S$ where $L$ is the horizon length, and under the imposition $%
S\propto A\propto L^{2}$ \cite{ref14}, the energy density of the Barrow
holographic dark energy is expressed as 
\begin{equation}
\rho _{B}=CL^{\Delta -2},  \label{e2}
\end{equation}%
where $C$ is a parameter with dimension $\left[ L\right] ^{-2-\Delta }$, $L$
can be considered as the size of the current Universe such as the Hubble's
scale and the future event horizon, and $\Delta $ is a free parameter. It
can be seen that $\rho _{B}$ reduces to the standard holographic dark energy
model $3M_{p}^{2}L^{-2}$ at $\Delta =0$, where $C=3M_{p}^{2}$. Generally, in
the case of the deformation effects produced by $\Delta $ when it is not
zero, we get the Barrow holographic dark energy that can be distinguished
from the standard expression. Recently, Nandhida and Mathew \cite{ref15}
considered the Barrow holographic dark energy as a dynamical vacuum, with
Granda-Oliveros length as IR cut-off \cite{Granda}\ and studied the
evolution of cosmological parameters with the best estimated model
parameters extracted using the combined data-set of SN-Ia and observational
Hubble's data (OHD). Adhikary et al. \cite{ref16} constructed a Barrow
holographic dark energy in the case of non-flat universe in particular,
considering closed and open spatial geometry and observed that the scenario
can describe the thermal history of the universe, with the sequence of
matter and dark energy epochs. Considering Barrow holographic dark energy
Sarkar and Chattopadhyay \cite{ref17} reconstruct $f(R)$ gravity as the form
of background evolution and point out the equation of state can have a
transition from quintessence to phantom with the possibility of Little Rip
singularity.

Another one way to look beyond the dark energy and its models is to change
the left hand side of the Einstein's equation, foremost to a modified
theories of gravity. A numerous modified theories of gravity are existing
such as $f(R)$ which is obtained by the replacement of Ricci scalar with an
arbitrary function of the Ricci scalar \cite{ref18, ref19, ref20}. Next, $%
f(T)$ is obtained by replacing torsion with curvature \cite{ref23, ref24,
ref25}. Next, $f(R,T)$ is obtained by the replacement of Ricci scalar with
an arbitrary function of both Ricci scalar and trace of energy tensor \cite%
{ref29, ref30, ref31, ref32}. Further, $f(G)$ in which $f(G)$ is a generic
function of the Gauss-Bonnet invariant $G$, \cite{ref37, ref38, ref39} have
taken remarkable effort in $f(G)$ gravity. Next, $f(R,G)$ which combines
Ricci scalar and Gauss-Bonnet scalar \cite{ref40, ref41}. After that, $%
f(T,B) $ which involve torsion and boundary term \cite{ref42}. Further,
recently proposed symmetric teleparallel gravity say $f(Q)$ gravity in which
non-metricity term $Q$ is responsible for the gravitational interaction \cite%
{ref43}. References \cite{Q1, Q2} include the first $f(Q)$ gravity
cosmological solutions, while \cite{Q3, Q4} provide the corresponding $f(Q)$
cosmography and energy conditions. A power-law model has been considered for
quantum cosmology \cite{Q5}. For a polynomial functional form of $f(Q)$,
cosmological solutions and growth index of matter perturbations have been
researched \cite{Q6}. By assuming a power-law function, Harko et al.
examined the coupling matter in $f(Q)$ gravity \cite{Q7}. An attractive
investigation of $f(Q)$ gravity was done by Frusciante, where he examined
the signatures of non-metricity gravity in its' fundamental level \cite{Q8}.

Hence, motivated with the above discussion and observations, in the present
article we studied Barrow holographic dark energy in the symmetric
teleparallel gravity with an anisotropic background. The out-lines of the
analysis are as: In Sec. \ref{sec2} we provide the brief review of $f(Q)$
gravity with the equation of source. In Sec. \ref{sec3}, we mention the
metric and components of field equations. The solution of the field
equations and the parameters of the cosmological models are found in Sec. %
\ref{sec4} while in Sec. \ref{sec5} an observational constraints from the
set of 57 Hubble's data point are obtained. Finally, we summarize our
results in conclusion Sec. \ref{sec6}.

\section{$f(Q)$ gravity and source}

\label{sec2}

In $f(Q)$ gravity, the action is given by \cite{ref43}

\begin{equation}  \label{e3}
S=\int \left[ \frac{1}{2}f(Q)+L_{m}\right] d^{4}x\sqrt{-g},
\end{equation}
where $f(Q)$ is an arbitrary function of the non-metricity $Q$, $g$ is the
determinant of the metric tensor $g_{\mu \nu }$ and $L_{m}$ is the matter
Lagrangian density. The non-metricity tensor and its two traces are such
that 
\begin{equation}  \label{e4}
Q_{\gamma \mu \nu }=\nabla _{\gamma }g_{\mu \nu },
\end{equation}
\begin{equation}
Q_{\gamma }=Q_{\gamma }{}^{\mu }{}_{\mu },\text{ \ \ \ \ }\overset{\sim }{Q}%
_{\gamma }=Q^{\mu }{}_{\gamma \mu }.  \label{e5}
\end{equation}

In addition, the superpotential as a function of non-metricity tensor is
given by 
\begin{equation}
P_{\mu \nu }^{\gamma }=\frac{1}{4}\left( -Q_{\mu \nu }^{\gamma }+2Q_{\left(
\mu ^{\gamma }\nu \right) }-Q^{\gamma }g_{\mu \nu }-\overset{\sim }{%
Q^{\gamma }}g_{\mu \nu }-\delta ^{\gamma }\left( _{\gamma }Q_{\nu }\right)
,\right)  \label{e6}
\end{equation}%
where the trace of non-metricity tensor \eqref{e5} has the form 
\begin{equation}
Q=-Q_{\gamma \mu \nu }P^{\gamma \mu \nu }.  \label{e7}
\end{equation}

In addition, we can also take the variation of \eqref{e3} with respect to
the connection, which gives 
\begin{equation}
\nabla _{\mu }\nabla _{\gamma }\left( \sqrt{-g}f_{Q}P^{\gamma }{}_{\mu \nu
}\right) =0.  \label{e8}
\end{equation}

The energy-momentum tensor is given by 
\begin{equation}
T_{\mu \nu }=-\frac{2}{\sqrt{-g}}\frac{\delta \left( \sqrt{-g}L_{m}\right) }{%
\partial g^{\mu \nu }}.  \label{e9}
\end{equation}

The variation of the action $\left( S\right) $ in Eq.\eqref{e3} with respect
to the metric, we get the gravitational field equation given by

\begin{equation}
\frac{2}{\sqrt{-g}}\nabla _{\gamma }\left( \sqrt{-g}f_{Q}P^{\gamma }{}_{\mu
\nu }\right) -\frac{1}{2}fg_{\mu \nu }+f_{Q}\left( P_{\mu \gamma i}Q_{\nu
}{}^{\gamma i}-2Q_{\gamma i\mu }P^{\gamma i}{}_{\nu }\right) =\left( T_{\mu
\nu }+\overline{T}_{\mu \nu }\right) ,  \label{e10}
\end{equation}%
where $f_{Q}=\frac{df}{dQ}$. The energy-momentum tensor for matter $\left(
T_{\mu \nu }\right) $ and Barrow holographic dark energy $(\overline{T}_{\mu
\nu })$ are defined as 
\begin{equation}
T_{\mu \nu }=\rho _{m}u_{\mu }u_{\nu }=diag\left[ -1,0,0,0\right] \rho _{m},
\label{e11}
\end{equation}%
and 
\begin{equation}
\overline{T}_{\mu \nu }=\left( \rho _{B}+p_{B}\right) u_{\mu }u_{\nu
}+g_{\mu \nu }p_{B}=diag\left[ -1,\omega _{B},\omega _{B},\omega _{B}\right]
\rho _{B},  \label{e12}
\end{equation}%
it can be parameterized as 
\begin{equation}
\overline{T}_{\mu \nu }=diag\left[ -1,\omega _{B},\left( \omega _{B}+\delta
\right) ,\left( \omega _{B}+\delta \right) \right] \rho _{B},  \label{e13}
\end{equation}%
where $\rho _{m}$, $\rho _{B}$ are the energy densities of matter and the
Barrow holographic dark energy, respectively, and $p_{B}$ is the pressure of
the Barrow holographic dark energy. Also, the equation of state parameter is
given by $\omega _{B}=\frac{P_{B}}{\rho _{B}}$ and $\delta $ in the above
equation is called the skewness parameter, representing the deviations from
equation of state parameter in $y$ and $z$ directions.

\section{Metric and symmetric teleparallel field equations}

\label{sec3}

We examine the spatially homogeneous and anisotropic Bianchi type-I
space-time mentioned by the following metric 
\begin{equation}
ds^{2}=-dt^{2}+A^{2}\left( t\right) dx^{2}+B^{2}\left( t\right) \left(
dy^{2}+dz^{2}\right) ,  \label{e14}
\end{equation}%
where $A\left( t\right) $ and $B\left( t\right) $ represents the metric
potentials of the Universe.\newline

The spatial volume $\left( V\right) $ and scale factor $\left( a\right) $ of
the space-time \eqref{e1} is, 
\begin{equation}
V=AB^{2}=a^{3},  \label{e15}
\end{equation}

The directional Hubble's parameters of the space-time \eqref{e1} is $H_{i}$ $%
\left( i=x,y,z\right) $, 
\begin{equation}
H_{x}=\frac{\overset{.}{A}}{A},\text{ \ \ \ \ }H_{y}=H_{z}=\frac{\overset{.}{%
B}}{B}.  \label{e16}
\end{equation}

Also, we assume that the deceleration parameter $q$ is a function of the
Hubble parameter $H$ as \cite{ref48} 
\begin{equation}
q=-\frac{a\overset{..}{a}}{\overset{.}{a}^{2}}=b-\frac{n}{H},  \label{e17}
\end{equation}%
where, $b$ and $n$ are constants, and $n>0$.

The motivation for presumed a variable deceleration parameter is the fact
that the expansion of the universe transits from the initial deceleration
phase to the current accelerated phase, confirmed by recent astrophysical
observations such as supernova type-Ia and cosmic microwave background (CMB)
anisotropy \cite{ref1, ref2, ref3, ref4}. The deceleration parameter is a
dimensionless variable quantity, given in Eq. \eqref{e17}. It describes the
cosmic evolution, if $q>0$ indicates the cosmic deceleration while $q<0$
indicates the acceleration of the expansion of the Universe.

The non-metricity scalar $Q$ for Bianchi type-I space-time is given by 
\begin{equation}
Q=-2\left( \frac{\overset{.}{B}}{B}\right) ^{2}-4\frac{\overset{.}{A}}{A}%
\frac{\overset{.}{B}}{B}.  \label{e18}
\end{equation}

The symmetric teleparallel field equations for Bianchi type-I space-time %
\eqref{e14} with the help of Eq. \eqref{e13} can be formulated as \cite%
{ref49}

\begin{equation}
\frac{f}{2}+f_{Q}\left[ 4\frac{\overset{.}{A}}{A}\frac{\overset{.}{B}}{B}%
+2\left( \frac{\overset{.}{B}}{B}\right) ^{2}\right] =\rho _{m}+\rho _{B},
\label{e19}
\end{equation}%
\begin{equation}
\frac{f}{2}-f_{Q}\left[ -2\frac{\overset{.}{A}}{A}\frac{\overset{.}{B}}{B}-2%
\frac{\overset{..}{B}}{B}-2\left( \frac{\overset{.}{B}}{B}\right) ^{2}\right]
+2\frac{\overset{.}{B}}{B}\overset{.}{Q}f_{QQ}=-\omega _{B}\rho _{B},
\label{e20}
\end{equation}%
\begin{equation}
\frac{f}{2}-f_{Q}\left[ -3\frac{\overset{.}{A}}{A}\frac{\overset{.}{B}}{B}-%
\frac{\overset{..}{A}}{A}-\frac{\overset{..}{B}}{B}-\left( \frac{\overset{.}{%
B}}{B}\right) ^{2}\right] +\left( \frac{\overset{.}{A}}{A}+\frac{\overset{.}{%
B}}{B}\right) \overset{.}{Q}f_{QQ}=-\left( \omega _{B}+\delta \right) \rho
_{B},  \label{e21}
\end{equation}%
where an over dot $\left( \overset{.}{}\right) $\ denote derivatives with
respect to the cosmic time $t$. In the coming sections, we will find the
exact solutions to the field equations \eqref{e19}-\eqref{e21}, so we will
give expressions for some of the quantities that we will need in this
context.

\section{Solution of field equations and cosmological models}

\label{sec4}

Now, we only get three independent field equations with seven unknown
parameters, i.e. $A$, $B$, $\rho _{m}$, $\rho _{B}$, $\omega _{B}$, $\delta $%
, $f$. Therefore, the system of equations \eqref{e19} - \eqref{e21} is
undetermined and supplementary equations relating these parameters are
needed to obtain explicit solutions of this system. In the literature, the
various investigators use several different assumptions that can be made to
solve this system: the shear scalar $\left( \sigma ^{2}\right) $\ in the
model is proportional to the scalar expansion $\left( \theta \right) $,
which leads to a relationship between metric potentials as 
\begin{equation}
A=B^{m},  \label{e22}
\end{equation}%
where $m\neq 1$ is a positive constant which takes care of the anisotropy
behavior of the space-time. The solution of the field equations is obtained
by considering the quadratic form of the $f\left( Q\right) $\ function as 
\cite{Qua1, Qua2, Q3} 
\begin{equation}
f\left( Q\right) =\lambda Q^{2},  \label{e23}
\end{equation}%
where $\lambda <0$ is a constant. Solving Eq. \eqref{e17} we obtain the
following form for the scale factor 
\begin{equation}
a=k\left( e^{nt}-1\right) ^{\frac{1}{1+b}},  \label{e24}
\end{equation}%
where $k$ is a constant. The following equation accounts for the
relationship between the scale factor $a$ and red-shift $z$ 
\begin{equation}
a=\frac{a_{0}}{1+z},  \label{e25}
\end{equation}%
where $a_{0}$ is the current value of the scale factor. Here, we take $%
a_{0}=1$.\newline

The above Eq. \eqref{e24} can be reformulated as 
\begin{equation}
\frac{1}{1+z}=k\left( e^{nt}-1\right) ^{\frac{1}{1+b}}.  \label{e26}
\end{equation}

From Eq. \eqref{e26}, the expression for cosmic time $t$ in terms of
red-shift $z$ is obtained as 
\begin{equation}
t\left( z\right) =\frac{1}{n}\log \left[ 1+\frac{1}{\left( k\left(
1+z\right) \right) ^{1+b}}\right] .  \label{e27}
\end{equation}

Now, from Eqs. \eqref{e15} and \eqref{e23}, we get the metric potentials 
\begin{equation}
A=k_{1}\left( e^{nt}-1\right) ^{\frac{3m}{\left( 1+b\right) \left(
2+m\right) }},  \label{e28}
\end{equation}%
\begin{equation}
B=C=k_{2}\left( e^{nt}-1\right) ^{\frac{3}{\left( 1+b\right) \left(
2+m\right) }}.  \label{e29}
\end{equation}%
where, $k_{1}=k^{\frac{3m}{2+m}}$\ and $k_{2}=k^{\frac{3}{2+m}}$.\

Using the metric potentials \eqref{e28} and \eqref{e29} in space-time %
\eqref{e14}, we can write 
\begin{widetext}
\begin{equation}
ds^{2}=-dt^{2}+k_{1}^{2}\left( e^{nt}-1\right) ^{\frac{6m}{\left( 1+b\right)
\left( 2+m\right) }}dx^{2}+k_{2}^{2}\left( e^{nt}-1\right) ^{\frac{6}{\left(
1+b\right) \left( 2+m\right) }}\left( dy^{2}+dz^{2}\right) .  \label{e30}
\end{equation}
\end{widetext}

Using Eq. \eqref{e24}, the mean Hubble's parameter $H$ can be derived as 
\begin{equation}
H=\frac{\overset{.}{a}}{a}=\frac{1}{3}\left( H_{1}+2H_{2}\right) =\frac{%
ne^{nt}}{\left( 1+b\right) \left( e^{nt}-1\right) }.  \label{e31}
\end{equation}

From Eq. \eqref{e17}, the deceleration parameter is obtained as 
\begin{equation}
q=-1+\left( 1+b\right) e^{-nt},  \label{e32}
\end{equation}

\begin{figure}[tbp]
\centerline{\includegraphics[scale=0.68]{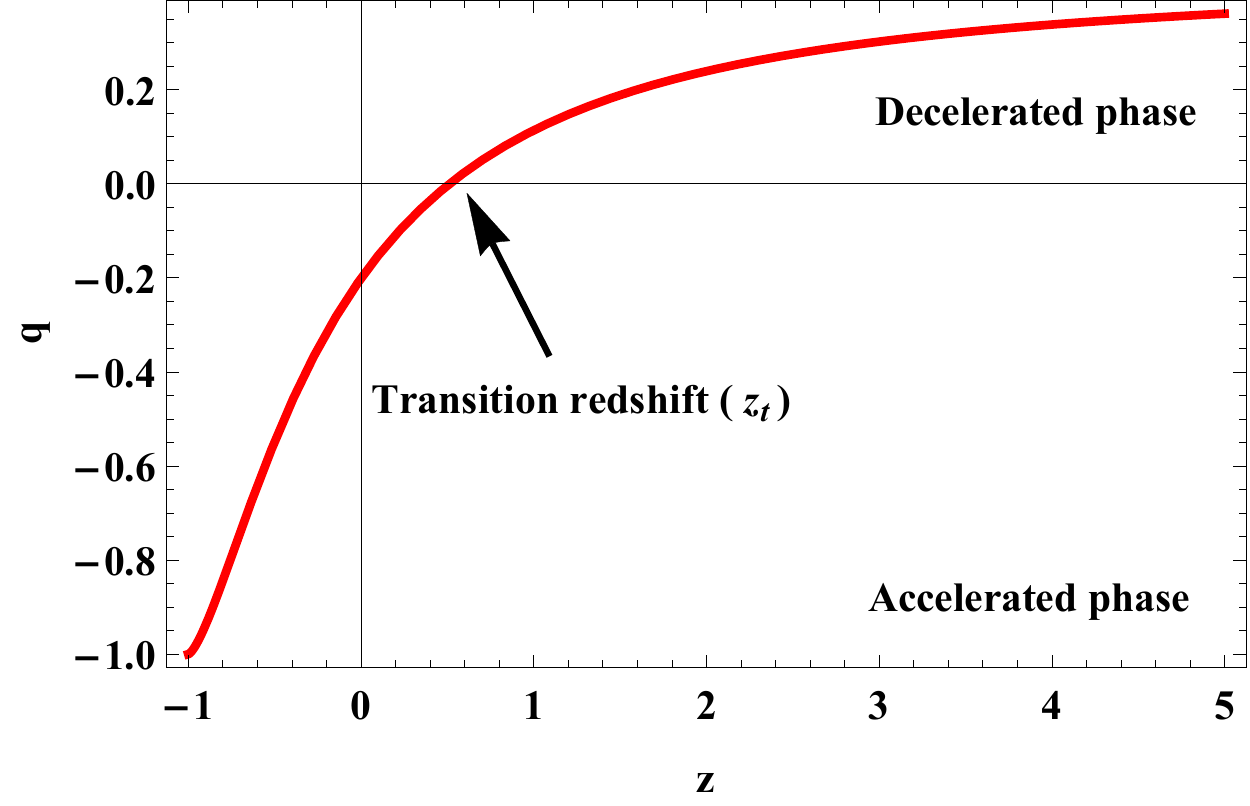}}
\caption{\emph{Plot of Deceleration parameter $\left( q\right) $ versus
red-shift $\left( z\right) $.}}
\label{DP}
\end{figure}

The model parameters $n$, $b$, and $k$ are constrained by Hubble datasets
(see Sec. \ref{sec5}). Also, the plot of the deceleration parameter versus
red-shift $\left( z\right) $ is shown in the Fig. \ref{DP}. It can be
observed from the Fig. \ref{DP} that the $q$ passes from positive to
negative value as the red-shift increases and it converges towards $-1$ when 
$z=-1$. Thus, our model of the Universe goes from an early deceleration
phase $\left( q>0\right) $ to a current acceleration phase $\left(
q<0\right) $. From Eq. \eqref{e32}, the phase transition occurs for $%
b=e^{nt}-1$ which leads to $q=0$. Therefore, our model is in good agreement
with recent observation data.

Using Eqs. \eqref{e28} and \eqref{e29} in \eqref{e18}, we get the
non-metricity scalar $Q$ as 
\begin{equation}
Q=\frac{-18\left( 1+2m\right) }{\left( m+2\right) ^{2}}H^{2}.  \label{e33}
\end{equation}

We examine the Hubble's horizon ( $L=H^{-1}$) as the IR cut-off of the
system, where $H$ is the Hubble's parameter of the model. Therefore, in
symmetric teleparallel gravity the energy density of Barrow holographic dark
energy \eqref{e2} takes the form 
\begin{equation}
\rho _{B}=CH^{2-\Delta }.  \label{e34}
\end{equation}

The motives for using this new formula introduced by Barrow is that he
thinks it is important in comprehension the evolution of the Universe for a
large horizon length $L$, in particular an anisotropic Universe.

From Eqs. \eqref{e31} and \eqref{e34} , we get the energy density of the
Barrow holographic dark energy as 
\begin{equation}
\rho _{B}=C\left[ \frac{ne^{nt}}{\left( 1+b\right) \left( e^{nt}-1\right) }%
\right] ^{2-\Delta }.  \label{e35}
\end{equation}

Using (\ref{e33}) and (\ref{e34}) in (\ref{e19}), the energy density of the matter in terms of Hubble's
parameter as 
\begin{equation}
\rho _{m}=\frac{-486\lambda \left( 1+2m\right) ^{2}}{\left( m+2\right) ^{4}}%
H^{4}-CH^{2-\Delta }.  \label{e36}
\end{equation}%
\begin{figure*}[tbp]
\begin{minipage}{0.45\linewidth}
\centerline{\includegraphics[scale=0.65]{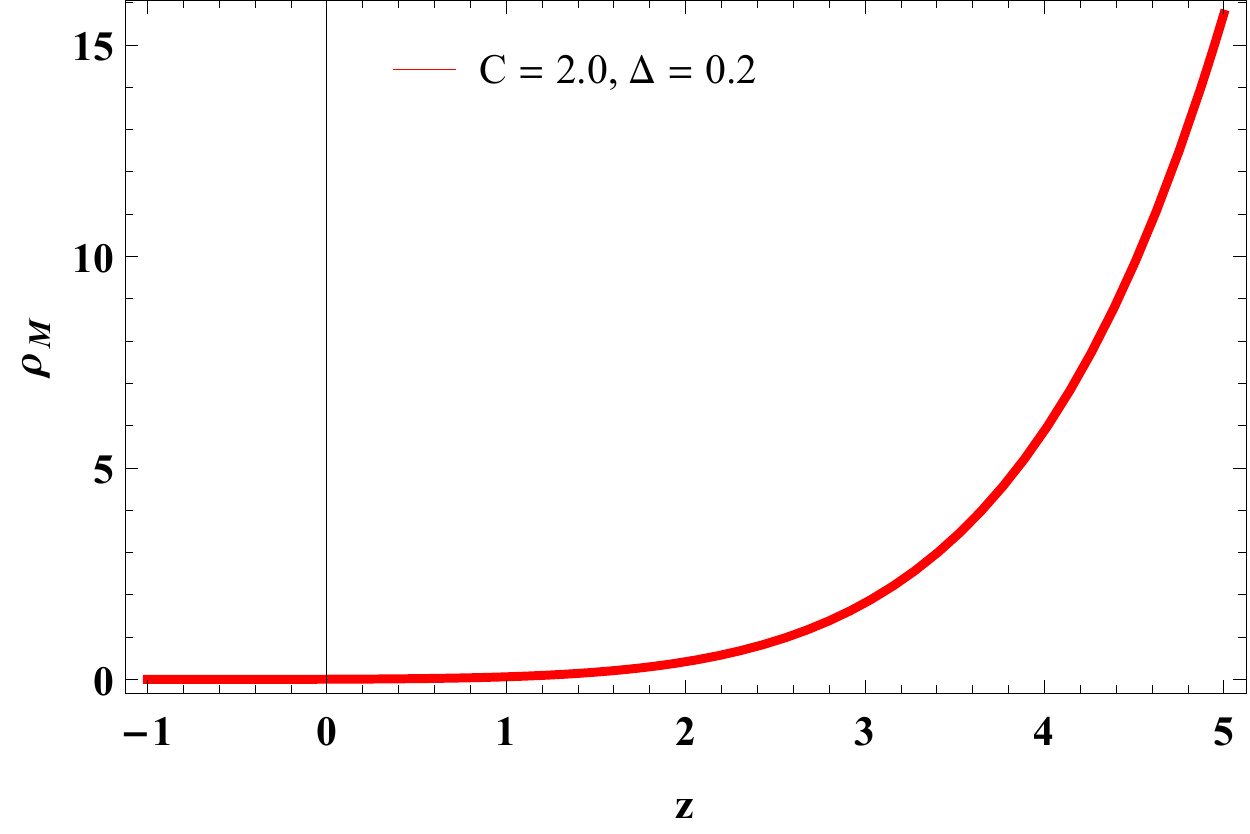}}
\caption{{\emph{Plot of energy density $\left( \protect\rho _{m}\right) $ of matter versus red-shift $\left( z\right)$.}}}\label{rhoM}
\end{minipage}
\hfill 
\begin{minipage}{0.45\linewidth}
\centerline{\includegraphics[scale=0.65]{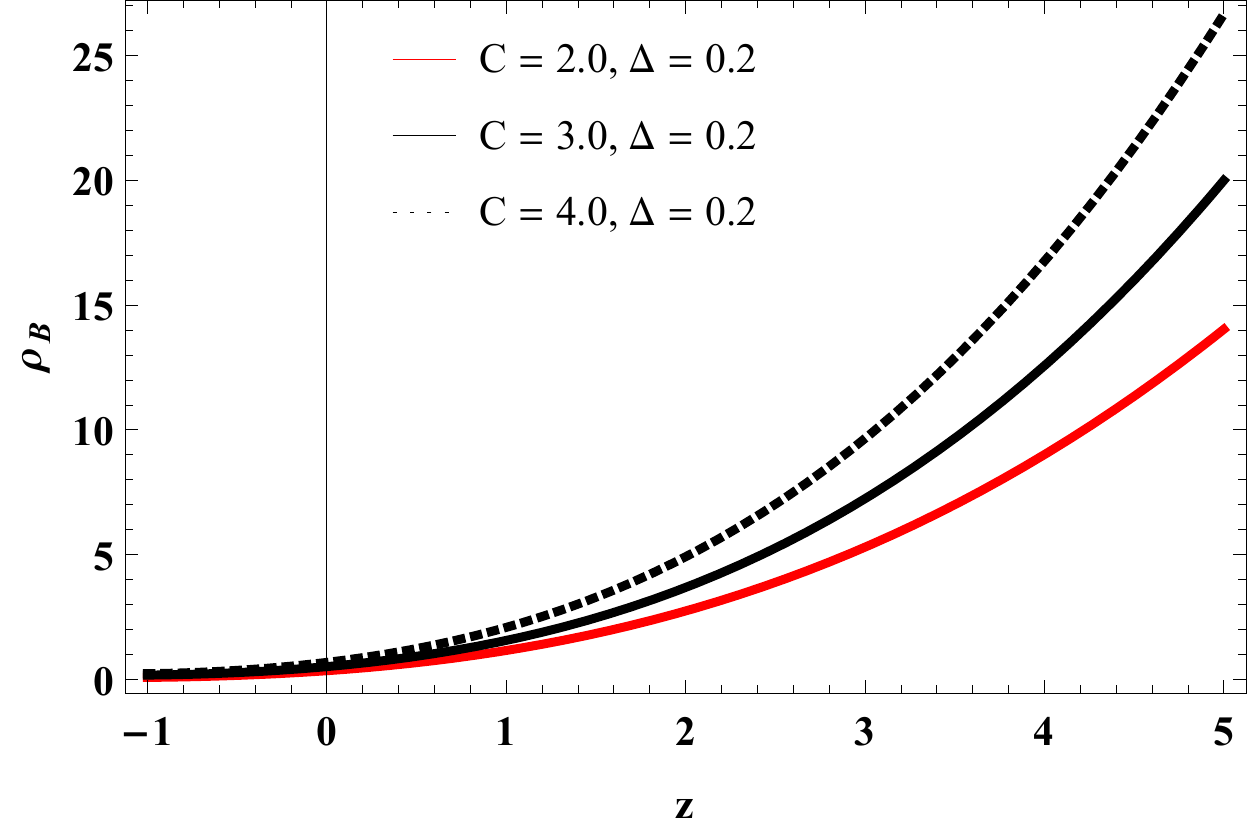}}
\caption{{\emph{Plot of energy density $\left( \protect\rho _{B}\right) $ of Barrow holographic dark energy versus red-shift $\left( z\right) $.}}}\label{rhoB}
\end{minipage}
\end{figure*}

In Figs. \ref{rhoM} and \ref{rhoB}, we plot the behavior of the energy
density of matter $\left( \rho _{m}\right) $ and Barrow holographic dark
energy $\left( \rho _{B}\right) $ with the Hubble's horizon cut-off with
respect to the red-shift $\left( z\right) $ for the appropriate values of
model parameters, respectively. It is observed that both $\rho _{m}$ and $%
\rho _{B}$ are positive and decreasing function of red-shift while energy
density of matter becomes null and energy density of Barrow holographic dark
energy attends a specific small constant value.

The skewness parameter $\delta $ as 
\begin{equation}
\delta =\frac{324\lambda \left( 1+2m\right) \left( 1-m\right) }{\rho
_{B}\left( m+2\right) ^{3}}H^{4}\left[ \left( 1+b\right) e^{-nt}-1\right] .
\label{e37}
\end{equation}

\begin{figure*}[tbp]
\begin{minipage}{0.45\linewidth}
\centerline{\includegraphics[scale=0.65]{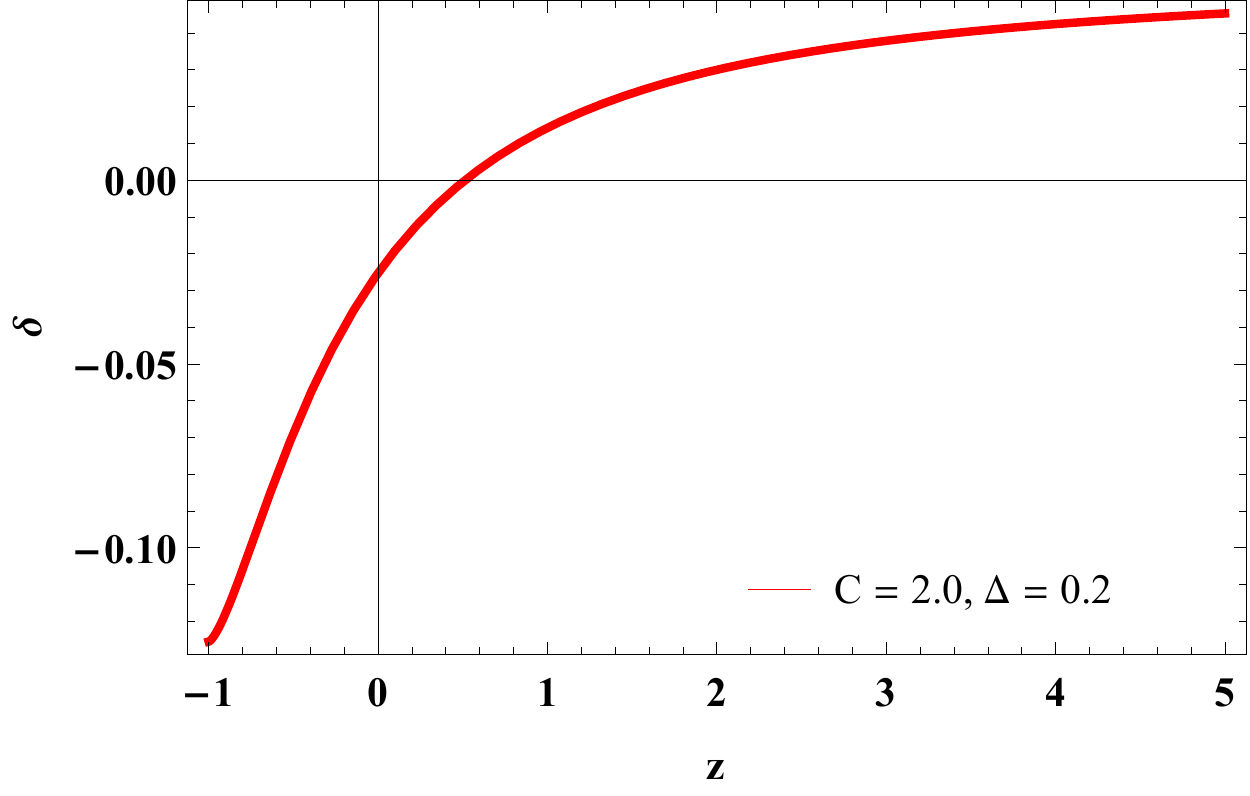}}
\caption{{\emph{Plot of skewness parameter $\left( \protect\delta \right) $
versus red-shift $\left( z\right) $}}}\label{sk}
\end{minipage}
\hfill 
\begin{minipage}{0.45\linewidth}
\centerline{\includegraphics[scale=0.65]{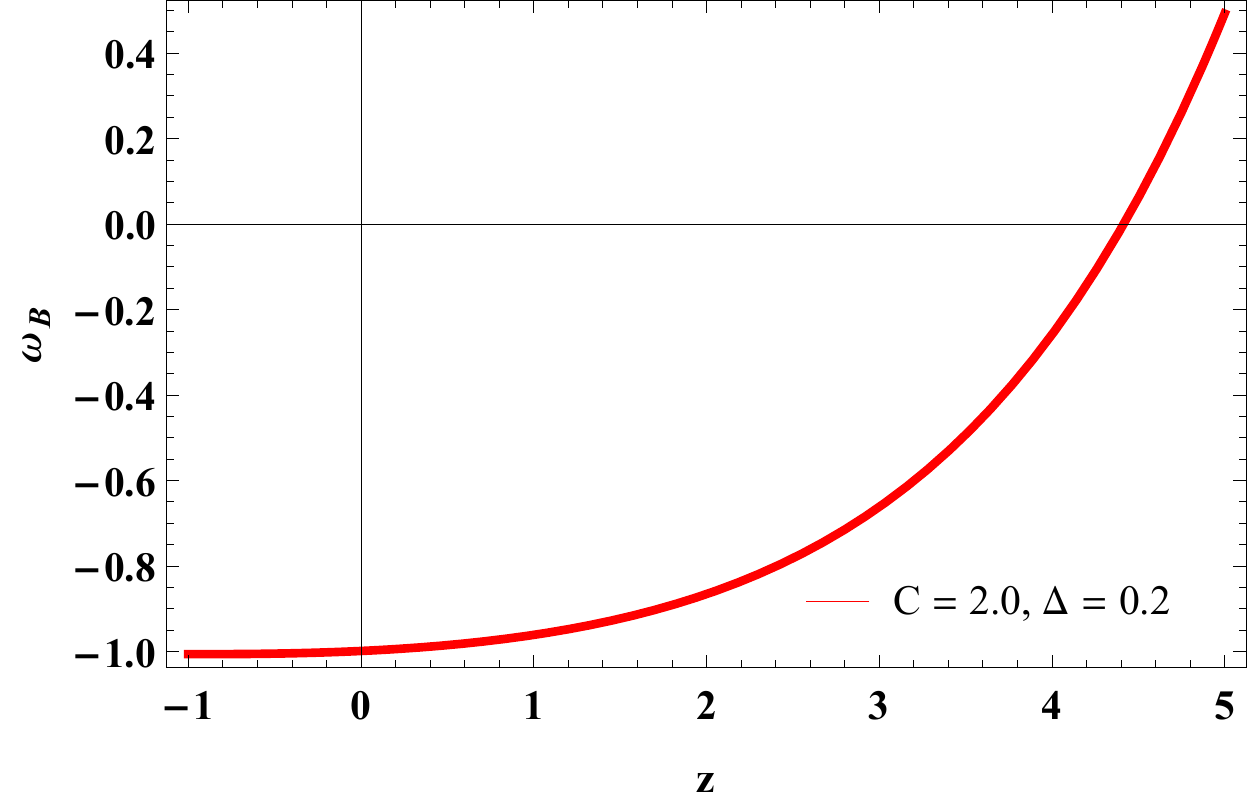}}
\caption{{\emph{Plot of $\left( \protect\omega _{B}\right) $ versus red-shift $\left( z\right) $.}}}\label{EoS}
\end{minipage}
\end{figure*}

Fig. \ref{sk} shows the evolution of the skewness parameter $\left( \delta
\right) $ versus red-shift $z$. It can be observed that $\delta $ of our
model varies between positive and negative values throughout cosmic
evolution. Moreover, we can see that the skewness parameter at the initial
epoch decreases until it reaches a negative value in the present epoch (i.e $%
z=0$) and in the future (i.e $z\rightarrow -1$). Therefore, our model is
anisotropic throughout the evolution of the Universe. From Eqs. \eqref{e35}
and \eqref{e36}, we observe that the energy densities of matter and Barrow
holographic dark energy are decreasing functions of cosmic time.

\begin{equation}
\rho _{m}=\frac{-486\lambda \left( 1+2m\right) ^{2}}{\left( m+2\right) ^{4}}%
H^{4}-CH^{2-\Delta }.  \label{e36}
\end{equation}%

The equation of state parameter of Barrow holographic dark energy is given
by 
\begin{widetext}
\begin{equation}
\omega _{B}=\frac{-162\lambda \left( 1+2m\right) }{C\left( m+2\right) ^{3}}%
H^{2+\Delta }\left[ \frac{\left( 1+2m\right) }{\left( m+2\right) }+4\left\{
\left( 1+b\right) e^{-nt}-1\right\} \right] .  \label{e38}
\end{equation}
\end{widetext}The evolution of equation of state is shown as a function of
red-shift $z$ in Fig. \ref{EoS}. From this figure, the equation of state
parameter of our model transitions from positive to negative value with
cosmic evolution, indicating the Universe's initial phase of deceleration
with positive pressure and the current phase of acceleration with negative
pressure. Also, at $z=0$, $\omega _{B}$ approaches $-1$ in the current era. 
\begin{figure*}[tbp]
\begin{minipage}{0.45\linewidth}
\centerline{\includegraphics[scale=0.65]{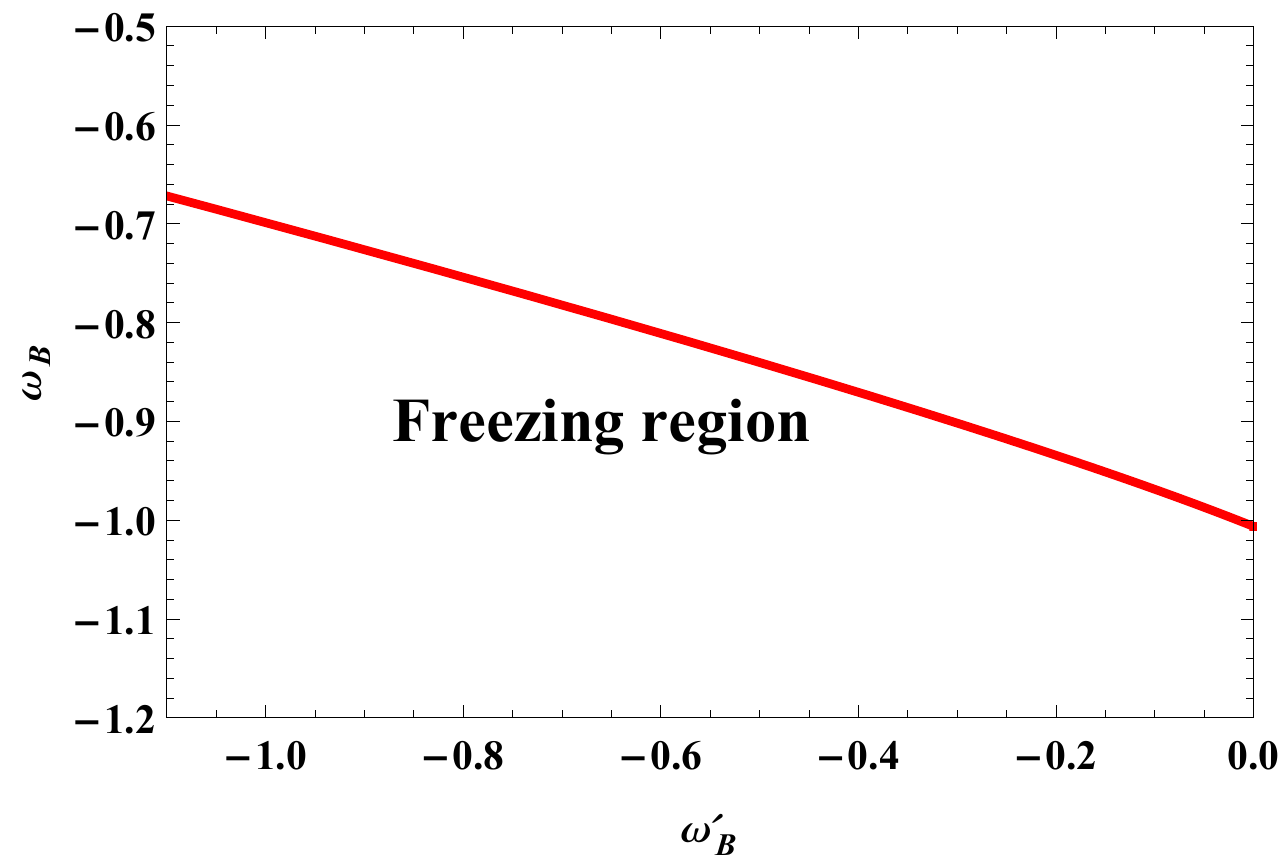}}
\caption{{\emph{Plot of $\protect\omega_{B}$ versus $\protect\omega _{B}^{^{\prime }}$.}}}\label{EoSP}
\end{minipage}
\hfill 
\begin{minipage}{0.45\linewidth}
\centerline{\includegraphics[scale=0.65]{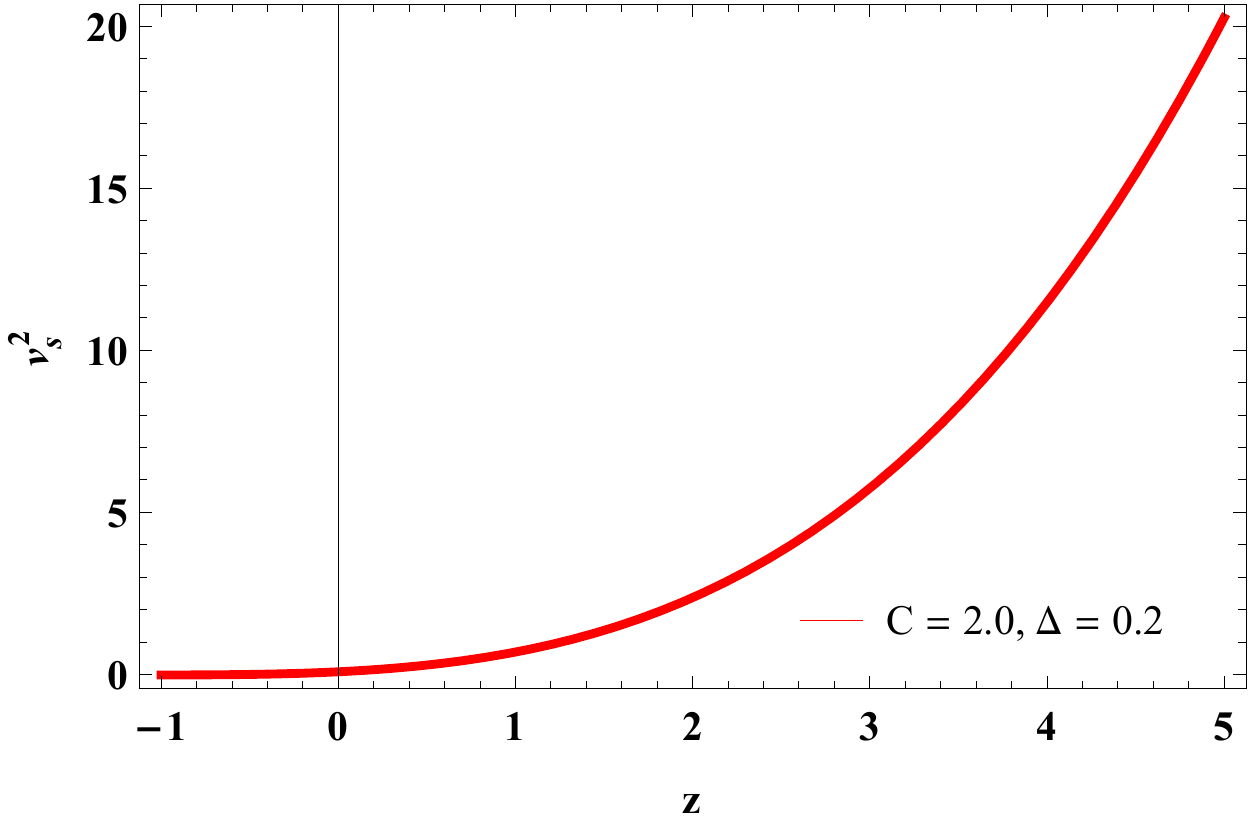}}
\caption{{\emph{Plot of squared sound speed $\left( v_{s}^{2}\right) $
versus red-shift $\left( z\right) $.}}}
\label{vs}
\end{minipage}
\end{figure*}

We can see from Eq.\eqref{e38}, that the equation of state parameter is a
function of time. However, the model starts from matter-dominated era,
varies in the quintessence region $-1<\omega _{B}<-\frac{1}{3}$\ and finally
approached to phantom region $\omega _{B}<-1$. At present the current value
of equation of state parameter is $\omega _{0}\sim -1$, i.e. the Universe is
dominated by $\Lambda $ cold dark matter $\left( \Lambda CDM\right) $. The
results obtained in our model are consistent with the constraints of the
equation of state parameter for the Planks Collaboration and WMAP which give
the ranges for Equation of state parameter as: $-0.92\leq \omega \leq -1.26$
(Planck+WP+Union 2.1), $-0.89\leq \omega \leq -1.38$ (Planck+WP+BAO), $%
-0.983\leq \omega \leq -1.162$\ (WMAP+eCMB+BAO+H0). Finally, the behavior of
the obtained model is in good agreement with recent observational data of
SNe-Ia. \newline

Next, the author Caldwell and Linder \cite{ref50}, proposed the ($\omega
_{B}-\omega _{B}^{^{\prime }}$)- plane to describe the dynamical properties
of the quintessence scalar field. In $\omega _{B}^{^{\prime }}$, prime
designate the derivative of equation of state parameter with respect to $%
x=\ln a$. By ($\omega _{B}-\omega _{B}^{^{\prime }}$)- plane analysis, we
consider two distinct thawing and freezing regions respectively corrosponds
to $\left( \omega _{B}^{^{\prime }}>0\text{, }\omega _{B}<0\right) $ and $%
\left( \omega _{B}^{^{\prime }}<0\text{, }\omega _{B}<0\right) $.\newline
Using Eq. \eqref{e38}, we get 
\begin{widetext}
\begin{equation}
\omega _{B}^{^{\prime }}=\frac{-162\lambda \left( 1+2m\right) }{C\left(
m+2\right) ^{3}}H^{1+\Delta }\left[ \left( 2+\Delta \right) \frac{H}{\overset%
{.}{H}}\left\{ \frac{\left( 1+2m\right) }{\left( m+2\right) }+4\left\{
\left( 1+b\right) e^{-nt}-1\right\} \right\} -4n\left( 1+b\right) e^{-nt}%
\right] .  \label{e39}
\end{equation}%
\end{widetext}where $\overset{.}{H}=\frac{-n^{2}e^{nt}}{\left(
e^{nt}-1\right) ^{2}\left( b+1\right) }$ .\newline
The plot of the ($\omega _{B}-\omega _{B}^{^{\prime }}$)- plane of the model
is given in Fig. \ref{EoSP}. Also, we can see that $\omega _{B}^{^{\prime
}}<0$\ and $\omega _{B}<0$ which indicates the freezing region of the
Universe. In this scenario, the squared sound speed $\left( v_{s}^{2}\right) 
$ is used for studying the stability of the dark energy models which is
given as $v_{s}^{2}=\frac{dp_{B}}{d\rho _{B}}=\frac{\overset{.}{p}_{B}}{%
\overset{.}{\rho }_{B}}$. If $v_{s}^{2}>0$, we obtain a stable model and if $%
v_{s}^{2}<0$, we obtain unstable model. For our Barrow holographic dark
energy model, $v_{s}^{2}$\ takes the following form 
\begin{widetext}
\begin{equation}
v_{s}^{2}=\frac{-648\lambda \left( 1+2m\right) }{C\left( m+2\right)
^{3}\left( 2-\Delta \right) }H^{2+\Delta }\left[ \frac{\left( 1+2m\right) }{%
\left( m+2\right) }+\left( 1+b\right) \left\{ 4-n\frac{H}{\overset{.}{H}}%
\right\} e^{-nt}-4\right]  \label{e40}
\end{equation}
\end{widetext}

Fig. \ref{vs} shows the evolution of the squared sound speed in terms of
red-shift $(z)$. It is clear that the Barrow holographic dark energy model
is stable i.e. $v_{s}^{2}>0$ with the cosmic expansion of the Universe.

Sahni et al. \cite{ref51} have instituted the geometrical diagnostic pair
named the state-finder pair $\left\{ r,s\right\} $ to differentiate dark
energy models. Here, $r$ is produced from the average scale factor $a$ and
its derivatives with regard to the cosmic time up to the third-order and $s$
is a simple composite of $r$. The state-finder pair $\left\{ r,s\right\} $
is defined as 
\begin{equation}
r=\frac{\overset{...}{a}}{aH^{3}}\text{, \ \ }s=\frac{r-1}{3\left( q-\frac{1%
}{2}\right) }.  \label{e41}
\end{equation}

The values of the state-finder parameter for our model are observed as 
\begin{equation}
r=1-\frac{3n+\left( 1+b\right) ^{2}\left( 1-e^{-2nt}\right) }{\left(
e^{nt}-1\right) }.  \label{e42}
\end{equation}%
\begin{equation}
s=\frac{2\left( 1+b\right) ^{2}\left( 1-e^{-2nt}\right) -6n}{\left(
e^{nt}-1\right) \left[ 6\left( 1+b\right) e^{-nt}-9\right] }.  \label{e43}
\end{equation}

\begin{figure}[tbp]
\centerline{\includegraphics[scale=0.65]{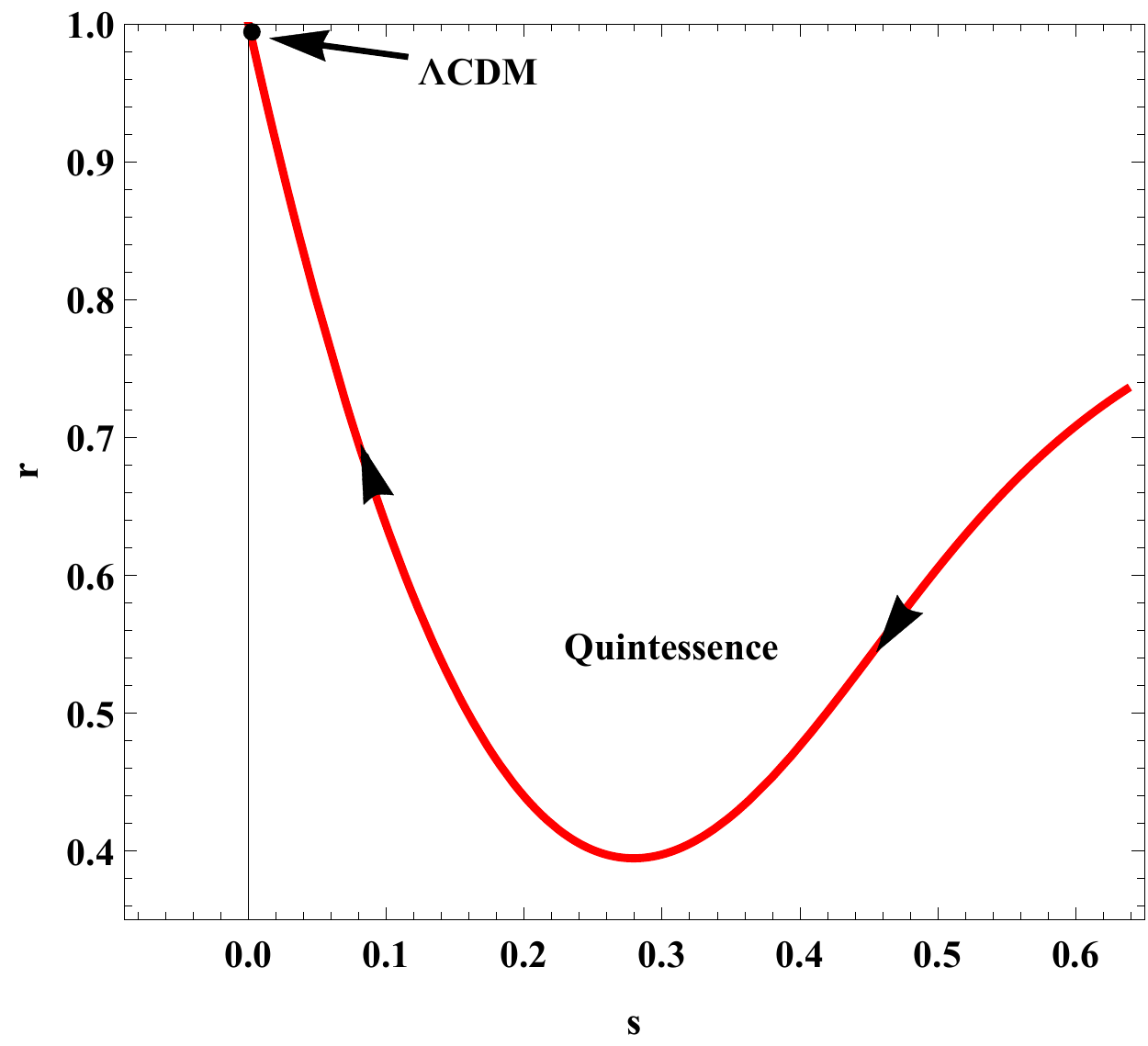}}
\caption{{\emph{Plot of state-finder parameter $\left\{ r,s\right\} $.}}}
\label{rs}
\end{figure}

For $\left\{ r=1,s=0\right\} $, we obtain the $\Lambda CDM$ model, while for 
$\left\{ r=1,s=1\right\} $, we obtain the cold dark matter $\left(
CDM\right) $ limit. In addition, for $\left\{ r<1,s>0\right\} $\ we obtain a
quintessence region. It can be observed from Fig. \ref{rs} that when $%
z\rightarrow -1$, $\left\{ r=1,s=0\right\} $ and as $z=0$, $\left\{
r<1,s>0\right\} $. This, indicates that our model starts from a quintessence
epoch and approaches the $\Lambda CDM$ Universe.

\section{Observational constraints}

\label{sec5}

In this section, we have constrained the model parameters with the
observational data. We have found the best fit value for the model
parameters $b$, $n$ and $k$ and the current value of the Hubble's parameter $%
H_{0}$ in our model. For this, we will use $51$ homogenized and
model-independent Hubble's parameter measurements in the range $0.07\leq
z\leq 2.36$ \cite{ref52}. In this set of $57$ Hubble's data points, $31$
points measured via the method of Differential Age (DA) and remaining $26$
points through BAO and other methods (see Tabl. \ref{Tab}) \cite{ref53}. The
Hubble's parameter $H$ in terms of red-shift $z$ as 
\begin{equation}
H\left( z\right) =\frac{n}{\left( 1+b\right) }\left[ 1+\left( k\left(
1+z\right) \right) ^{1+b}\right] .  \label{e44}
\end{equation}

The technique $R^{2}$-test to find the best fit value of the model
parameters defined by following statistical formula 
\begin{equation}
R^{2}=1-\frac{\sum_{i=1}^{57}\left[ \left( H_{i}\right) _{ob}-\left(
H_{i}\right) _{th}\right] ^{2}}{\sum_{i=1}^{57}\left[ \left( H_{i}\right)
_{ob}-\left( H_{i}\right) _{mean}\right] ^{2}},  \label{e45}
\end{equation}%
where, $\left( H_{i}\right) _{ob}$ and $\left( H_{i}\right) _{th}$ are
observed\ and predicted values of Hubble's parameter, respectively. 
\begin{figure}[h]
\centerline{\includegraphics[scale=0.5]{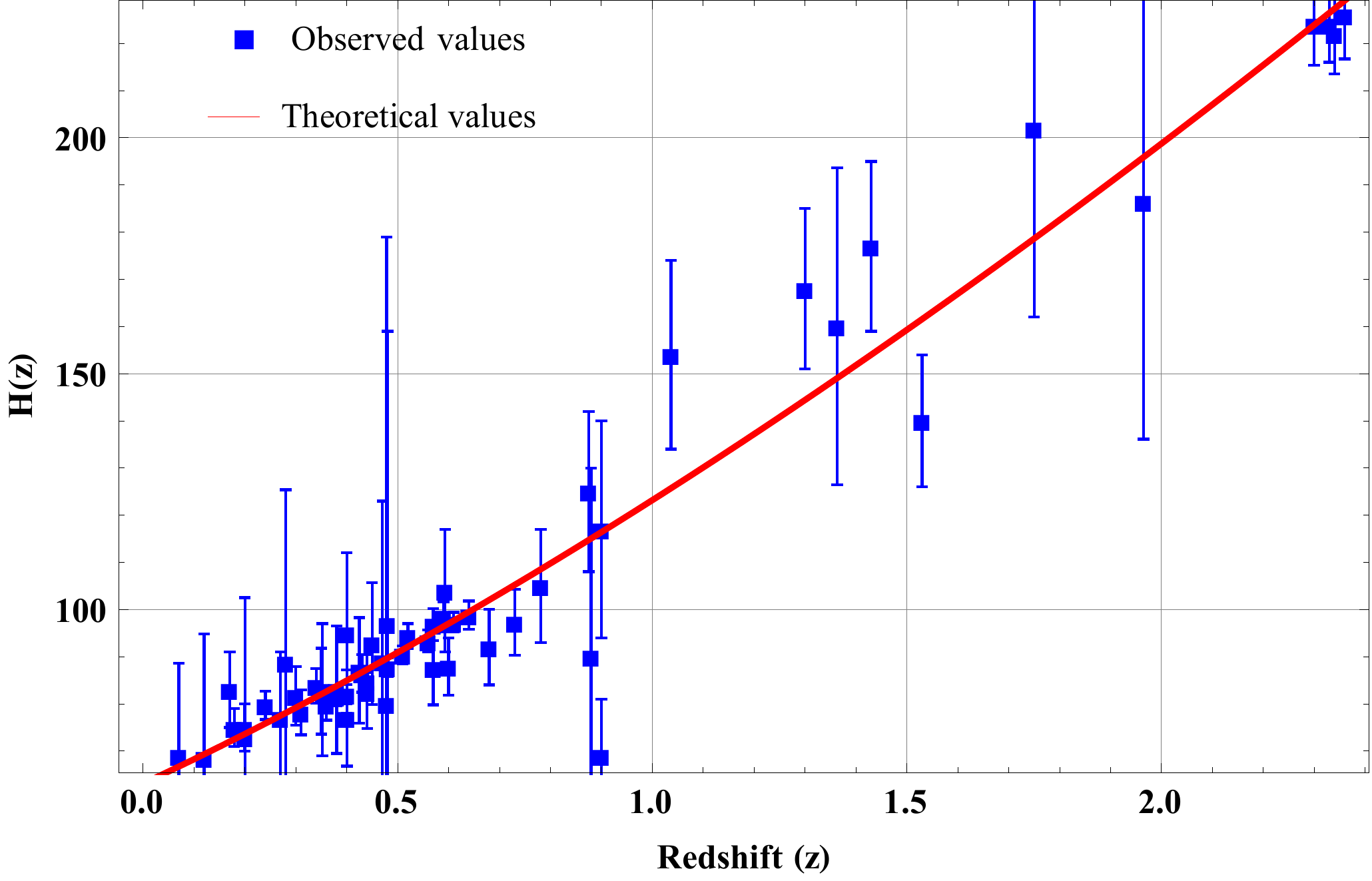}}
\caption{{\emph{Best fit curve of Hubble's parameter $\left( H\right) $
versus red-shift $\left( z\right) $.}}}
\label{Hubble}
\end{figure}

By minimizing $R^{2}$, we find the best fit values for the model parameters $%
b=0.444$, $n=40.73$ and $k=1.159$ (with 95\% confidence bounds). Also, we
find $R^{2}=0.9335$ and root mean square error $\left( RMSE\right) =11.54$
for the model under consideration with $57$ Hubble's parameter measurements.
If $R^{2}=1$ corresponds to the ideal case when the observed data and
corresponding values of theoretical function agree exactly. Now, Eq. %
\eqref{e44} becomes 
\begin{equation}
H\left( z\right) =28.2063+34.9047\left( 1+z\right) ^{1.444}.  \label{e46}
\end{equation}

Using Eq. \eqref{e46}, the current value of Hubble's parameter $H$ for the
obtained model, have calculated as $H_{0}=63.111\pm 11.54Km/s/Mpc$. Fig. \ref%
{Hubble} shows the best fit curve of the Hubble's parameter versus red-shift 
$z$ using $57$ Hubble's parameter measurements.

\begin{table}[tbp]
\begin{center}
\begin{tabular}{|c|c|c|c|c|c|}
\hline\hline
$z$ & $H(z)$ & $\sigma _{H}$ & $z$ & $H\left( z\right) $ & $\sigma _{H}$ \\ 
\hline\hline
0.070 & 69 & 19.6 & 0.4783 & 80 & 99 \\ \hline
0.90 & 69 & 12 & 0.480 & 97 & 62 \\ \hline
0.120 & 68.6 & 26.2 & 0.593 & 104 & 13 \\ \hline
0.170 & 83 & 8 & 0.6797 & 92 & 8 \\ \hline
0.1791 & 75 & 4 & 0.7812 & 105 & 12 \\ \hline
0.1993 & 75 & 5 & 0.8754 & 125 & 17 \\ \hline
0.200 & 72.9 & 29.6 & 0.880 & 90 & 40 \\ \hline
0.270 & 77 & 14 & 0.900 & 117 & 23 \\ \hline
0.280 & 88.8 & 36.6 & 1.037 & 154 & 20 \\ \hline
0.3519 & 83 & 14 & 1.300 & 168 & 17 \\ \hline
0.3802 & 83 & 13.5 & 1.363 & 160 & 33.6 \\ \hline
0.400 & 95 & 17 & 1.430 & 177 & 18 \\ \hline
0.4004 & 77 & 10.2 & 1.530 & 140 & 14 \\ \hline
0.4247 & 87.1 & 11.2 & 1.750 & 202 & 40 \\ \hline
0.4497 & 92.8 & 12.9 & 1.965 & 186.5 & 50.4 \\ \hline
0.470 & 89 & 34 &  &  &  \\ \hline\hline
$z$ & $H\left( z\right) $ & $\sigma _{H}$ & $z$ & $H\left( z\right) $ & $%
\sigma _{H}$ \\ \hline\hline
0.24 & 79.69 & 2.99 & 0.52 & 94.35 & 2.64 \\ \hline
0.30 & 81.7 & 6.22 & 0.56 & 93.34 & 2.3 \\ \hline
0.31 & 78.18 & 4.74 & 0.57 & 87.6 & 7.8 \\ \hline
0.34 & 83.8 & 3.66 & 0.57 & 96.8 & 3.4 \\ \hline
0.35 & 82.7 & 9.1 & 0.59 & 98.48 & 3.18 \\ \hline
0.36 & 79.94 & 3.38 & 0.60 & 87.9 & 6.1 \\ \hline
0.38 & 81.5 & 1.9 & 0.61 & 97.3 & 2.1 \\ \hline
0.40 & 82.04 & 2.03 & 0.64 & 98.82 & 2.98 \\ \hline
0.43 & 86.45 & 3.97 & 0.73 & 97.3 & 7.0 \\ \hline
0.44 & 82.6 & 7.8 & 2.30 & 224 & 8.6 \\ \hline
0.44 & 84.81 & 1.83 & 2.33 & 224 & 8 \\ \hline
0.48 & 87.90 & 2.03 & 2.34 & 222 & 8.5 \\ \hline
0.51 & 90.4 & 1.9 & 2.36 & 226 & 9.3 \\ \hline
\end{tabular}%
\end{center}
\caption{57 points of $H(z)$ data: $31$ (DA) + $26$ (BAO+other) \protect\cite%
{ref53}.}
\label{Tab}
\end{table}

\section{Conclusions}

\label{sec6}

In this paper, we have considered a spatially homogeneous and anisotropic
Bianchi type-I space-time in the presence of Barrow holographic dark energy
(infrared cut-off is the Hubble's horizon) suggested by Barrow recently
(Physics Letters B 808 (2020): 135643) and matter in quadratic form of $f(Q)$
gravity i.e. $f(Q)=\lambda Q^{2}$ (where $\lambda <0$ is a constant). The
exact solutions to the field equations are obtained assuming that the
deceleration parameter $q$ is a function of the Hubble's parameter $H$ i.e. $%
q=b-\frac{n}{H}$ (where $b$ and $n$ are constants). First, we have studied
the behavior of this form of deceleration parameter given in Eq. \eqref{e32}
and it turns out that passes from positive to negative value as the
red-shift increases and it converges towards $-1$ when $z=-1$. Hence, our
model of the Universe goes from an early deceleration phase to a current
acceleration phase which is in good agreement with recent observation data.
Next, we have got the energy densities of matter and Barrow holographic dark
energy both are positive decreases as Universe expands, the energy density
of matter becomes null and energy density of Barrow holographic dark energy
attends a specific small constant value. Also, the equation of state
parameter of our model transitions from positive to negative value with
cosmic evolution. The model starts from matter-dominated era, varies in the
quintessence region $-1<\omega _{B}<-\frac{1}{3}$\ and finally approached to 
$\Lambda $CDM region ($\omega _{B}=-1$). At present the current value of
equation of state parameter is $\omega _{0}\sim -1$, i.e. the Universe is
dominated by $\Lambda CDM$. Thus, the behavior of the obtained model is in
good agreement with recent observational data.

The skewness parameter represents same behavior as that of the equation of
state parameter. At the initial epoch it decreases until reaches a negative
value in the present and future epoch which confirms that the model is
anisotropic throughout the evolution of the Universe. For the proposed plane
of Caldwell and Linder, it is observed that the equation of state parameter
and the argument of equation of state parameter with lna both are
non-negative which represents freezing region of the Universe where as the
stability parameter in term of squared sound speed is stable i.e. $%
v_{s}^{2}>0$ with the cosmic expansion of the Universe. In $\left\{
r,s\right\} $-plane, Initially $\left\{ r<1,s>0\right\} $ and with cosmic
expansion $\left\{ r,s\right\} =\left\{ 1,0\right\} $. This, indicates that
our model starts from a quintessence epoch and approaches the $\Lambda CDM$
Universe.

\acknowledgments We are very much grateful to the honorary referee and the
editor for the illuminating suggestions that have significantly improved our
work in terms of research quality and presentation.\newline

\end{document}